\documentclass[twocolumn,showpacs,aps]{revtex4}
\usepackage[dvips]{graphicx}

\begin{document}

\title{Optimal quantum parameter estimation of two
interacting qubits under decoherence}


\author
{
Zheng Qiang $^{1,~2}$, Yao Yao $^{1}$,
Yong Li $^{1}$ 
}
\address{
$^{1}$ Beijing Computational Science Research Center, Beijing 100084, China
\\
$^{2}$ School of Mathematics and Computer Science, Guizhou Normal
University, Guiyang, 550001, China}

\begin{abstract}
We investigate the parameter estimation problem in a two-qubit system,
in which each qubit is independently interacting
with its Markovian environment.
We study in detail the sensitivity of the estimation on the decoherence rate $\gamma$
and the two-qubit interaction strength $v$.
In particular, the dynamics of quantum Fisher information are employed
as a measure to quantify the precision of the estimations.
We find that the quantum Fisher information with respect to
the decoherence rate scales like
$t^2/[\exp(\gamma t)-1]$ and $\exp(-2\gamma t)$ in the
unitary limit and the completely decoherent limit, respectively.
When we estimate the interaction strength $v$,
the quantum Fisher information shows oscillation behavior in term of time.
In addition, our results provide further evidence that the entanglement of
the input state may not enhance quantum metrology.
\end{abstract}


\maketitle

\section{Introduction}
Each quantum system is in contact with its environment.
Understanding the dynamics of physical quantities under the effect
of its decoherence environment has attracted much more interest.
On a foundational level, time is a basically physical quantity.
So the dynamical evolution is an important property of system,
which makes the finite-time
quantum quantities interesting in their own right \cite{Tsang13}.
An well-known example is that the
entanglement dynamics of two qubits exhhibits
entanglement sudden death \cite{Yu06} in the decoherent environments.

The environmental noise presented in the physical system
often determines the performance of quantum property.
Therefore, it is important to develop
methods to estimate the level of noise as precisely as possible.
Determining the environmental parameters affecting
a quantum system is the first step to develop means to
control its spoiling effects. Except quantum process tomography \cite{Nielsen00},
quantum channel estimation \cite{Fujiwara01}
is an efficient way to identify an unknown noise by estimating noise parameters.
A general description of quantum channel estimation is as follows: for a prepared state
$\rho$ as an input of quantum channel $\Upsilon_{\theta}$, the channel parameters may be
estimated efficiently by performing some quantum state measurements
on the output state $\Upsilon_{\theta}(\rho)$. Thus, one should seek an optimal input state
$\rho$ and/or an optimal measurement on the output state $\Upsilon_{\theta}(\rho)$. Very
recently, based on optical setup, the experimental realization of
optimal estimation is reported for a Pauli noisy channel \cite{Chiuri04}.

Fisher information \cite{Fisher} is a key quantity in classical
estimation theory. Extending to quantum regime, quantum Fisher information (QFI)
is also very important in quantum estimation theory and quantum information theory.
QFI characterizes the sensitivity
of a state with respect to changes on a parameter \cite{Caves94}.
It is also related to Cram$\acute{e}$r-Rao inequality \cite{Holevo},
which determines the bound of the optimal
measurement.
In the field of quantum estimation, the aim is to determine
the value of the unknown parameter labeling the quantum
system, and the primary goal is to enhance the precision of
resolution. Moreover, QFI is also closely related with other quantity,
especially entanglement \cite{Smerzi09}.

Previously, Huelga \textit{et.~al.} have discussed the
optimal precision of frequency measurements in the presence of decoherence \cite{Huelga07}.
In a recent paper, a general theory for the quantum metrology of noisy systems has been proposed by
Escher and co-workers \cite{Escher11}.
More recently, the dynamics of QFI under decoherence excites wide interest.
It has been shown that the evolution of QFI under decoherence for the $N$-qubit GHZ
state shows the decay and sudden change \cite{wangx11}.
For a spin-$j$ system surrounded by a quantum critical Ising chain,
the QFI decays almost monotonously when the environment reaches
the critical point \cite{zsun10}.
The dynamics of QFI for a qubit subject to a
non-Markovian environment shows revival and retardation loss \cite{Berrada}.

The main aim of this paper is to examine the problem of parameter estimation
for two initially entangled qubits under decoherence.
In particular, we focus on the dynamical evolution of quantum Fisher information.
Quantum Fisher information shows the revival-like behavior in the Markovian environment
when the estimated parameter is their interacting strength.
This generalizes the result in Ref.~\cite{Berrada}.
We also show that quantum Fisher information can not be enhanced by
the entanglement of initially mixed state.
This extends the general result that the entangled states can
improve the precision of parameter estimation \cite{Kok, Bollinger} from
the view point of QFI dynamics.

\section{Quantum Fisher information}
The classical Fisher information is defined as
\begin{equation}
\begin{array}{ll}
F_{\theta,~c}=\sum_{i}p_{i}(\theta)[\frac{\partial}{\partial \theta} \ln p_{i}(\theta)]^2,
\end{array} \label{classFish}
\end{equation}
where $p_{i}(\theta)$ is the probability density conditioned on the fixed parameter $\theta$
with measurement outcome $\{ x_{i} \}$ for a discrete observable X. The classical Fisher information characterizes the inverse variance of the asymptotic
normality of a maximum-likelihood estimator.

Extending to quantum regime, quantum Fisher
information of a parameterized quantum states $\rho(\theta)$ is defined as
\begin{equation}
\begin{array}{ll}
F_{\theta}= \mathrm{Tr} [\rho(\theta) L^2],
\end{array}
\end{equation}
where $\theta$ is the parameter to be measured, and $L$ is the symmetric
logarithmic derivative determined by
\begin{equation}
\begin{array}{llll}
\frac{d \rho(\theta)}{d {\theta} }=\frac{1}{2}[\rho(\theta) L+ L \rho(\theta)].
\end{array}
\end{equation}
With the spectrum decomposition $\rho_{\theta}= \sum_{k} \lambda_{k}|k\rangle \langle k|$,
its QFI with respect to $\theta$ is given as \cite{Zhong13, Berrada12, Berrada13b}
\begin{equation}
F_{\theta}= \sum_{k} \frac{(\partial_{\theta }\lambda_{k})^2}{\lambda_{k}}+
2\sum_{k, k'} \frac{(\lambda_{k}-\lambda_{k'})^2}{\lambda_{k}+\lambda_{k'}} |\langle k | \partial_{\theta}k' \rangle|^2.
\label{FisherA}
\end{equation}
Here $\lambda_{k}>0$ and $\lambda_{k}+\lambda_{k'}>0$.
The first term in Eq.~(\ref{FisherA}) is just the classical Fisher information Eq.~(\ref{classFish}).
Then, the second term can be considered as the quantum contribution. According to quantum Cram$\acute{e}$r-Rao (QCR) inequality, the variance Var($\hat{\theta}$) of any
unbiased estimator $\hat{\theta}$ satisfies
\begin{equation}
\begin{array}{ll}
 \mathrm{Var}(\hat{\theta}) \geq \frac{1}{M F_{\theta}},
\end{array}
\end{equation}
where $M$ is the times of measurement.
QCR embodies the ultimate limit to the precision of the estimate of
$\theta$. With a larger quantum Fisher information, the parameter $\theta$
can be estimated more accurately.

\begin{figure}[!tb]
\centering
\includegraphics[width=3.0in]{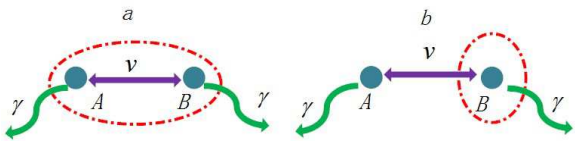}
\hspace{2.5cm}
\caption{ Schematic diagrams of the parameter estimation adopted in this paper.
Two qubits A and B couple to each other by the interaction parameter $v$. They also
have equally local decoherent rates $\gamma$.
(a) corresponds to the estimate of $\gamma$ taking the two qubits as a whole,
(b) corresponds to the estimate of $v$ only using qubit $B$.}
\label{system}
\end{figure}

\section{Quantum Fisher information of two qubits}
\subsection{Model}
In order to study the the precision of parameter estimation,
we consider a model consisting of two identically interacting
qubits A and B \cite{Agarwal09}. This simple model can make us get analytical results
for the time evolution of QFI.
Each qubit is a two-level system with an excited state $|e\rangle$
and a ground state $|g\rangle$. The interacting Hamiltonian is given by
\begin{equation}
\begin{array}{llll}
H = \frac{1}{2}v(S_{A}^{+}S_{B}^{-}+h.c.).
\end{array}\label{Hamit}
\end{equation}
Here $v$ denotes the interaction strength between the two qubits, $S_{j}^{+}$
and $S_{j}^{-}$ ($j$=A, B) are the qubit's raising and
lowering operators, respectively.
Furthermore, each qubit interacts independently with its Markovian environment.
The decoherence may arise from the spontaneous emission of the excited state.
The dynamics of the system can be treated by quantum Liouville equation ($\hbar=1$)
\begin{equation}
\frac{d \rho(t)}{d t}=-i[H, \rho]-\sum_{j=A, B}\frac{\gamma_{j}}{2}(S_{j}^{+}S_{j}^{-}\rho
-2S_{j}^{-}\rho S_{j}^{+} + \rho S_{j}^{+}S_{j}^{-}),
\label{liouv}
\end{equation}
where $\rho(t)$ is the density operator of the two qubits,
$\gamma_{j}$ are their spontaneous decay rates. The number of elements in the
density matrix $\rho(t)$ is $16$.


\begin{figure}[!tb]
\centering
\includegraphics[width=2.5in]{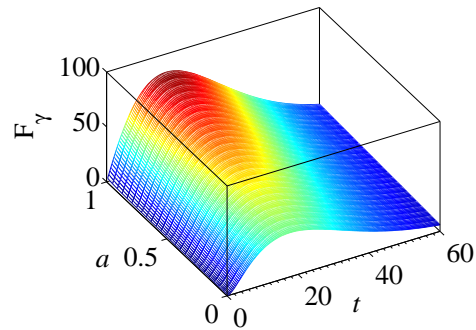}
\hspace{2.2cm}
\caption{The variation of $F_{\gamma}$ in Eq.~(\ref{FisherGamma}) with respect to the
time $t$ and the parameter $a$. Here the decoherence parameter $\gamma=0.1$.} \label{Fs1}
\end{figure}

In this paper, we consider a simple class of initial state
\begin{equation}
\rho_{0}=\frac{1}{3}(a|ee\rangle \langle ee|+d|gg\rangle \langle gg|+|\psi \rangle \langle \psi| )
=\frac{1}{3} \left(                 
\begin{array}{cccc}   
 a & 0 & 0& 0 \\  
 0 & 1 & z& 0 \\  
 0 & z^{*} & 1& 0 \\
 0 & 0 & 0& 1-a \\
 \end{array}
 \right), \label{init}
\end{equation}
with $|\psi \rangle= |eg\rangle+z |ge\rangle$, and the
parameters satisfying $0\leq a\leq 1$, $d$=1-$a$ and
$z=\exp(i\chi)$ \cite{Yu04}. This initial state is just the
input state for the parameter estimation. The concurrence of $\rho_{0}$ is
\begin{equation}
C(\rho_{0})=\frac{2}{3}[1-\sqrt{a(1-a)}].
\end{equation}
It is easy to prove that the solution of the
quantum Liouville equation Eq.~(\ref{liouv})
preserves the form in Eq.~(\ref{init}) all the time \cite{Agarwal09}.

After some straightforward calculations,
the nonzero elements of density matrix $\rho(t)$ are found as
\begin{equation}
\begin{array}{ll}
\rho_{11}=\frac{1}{3} a p(-t)^2,
\\
\rho_{22}=\frac{1}{3} p(-t)^2[ p(t)(a-\sin (\chi ) \sin (2 t v)+1)-a],
\\
\rho_{23}=\frac{1}{6} p(-t) z^{*}[ (-1+z^{2}) \cos (2 t
v)+z^{2}+1],
\\
\rho_{33}=\frac{1}{3} p(-t)^2 [ p(t)(a+\sin (\chi ) \sin (2 t v)+1)-a ],
\\
\rho_{32}=\rho_{23}^{*},
\rho_{44}=1-\rho_{11}-\rho_{22}-\rho_{33}.
\end{array} \label{dens}
\end{equation}
Here we have assumed that two qubits have equal decay
rates $\gamma_{A}$=$\gamma_{B}$=$\gamma$ and $p(t)\equiv e^{\gamma t}$.

\begin{figure}[!tb]
\centering
\includegraphics[width=1.5in]{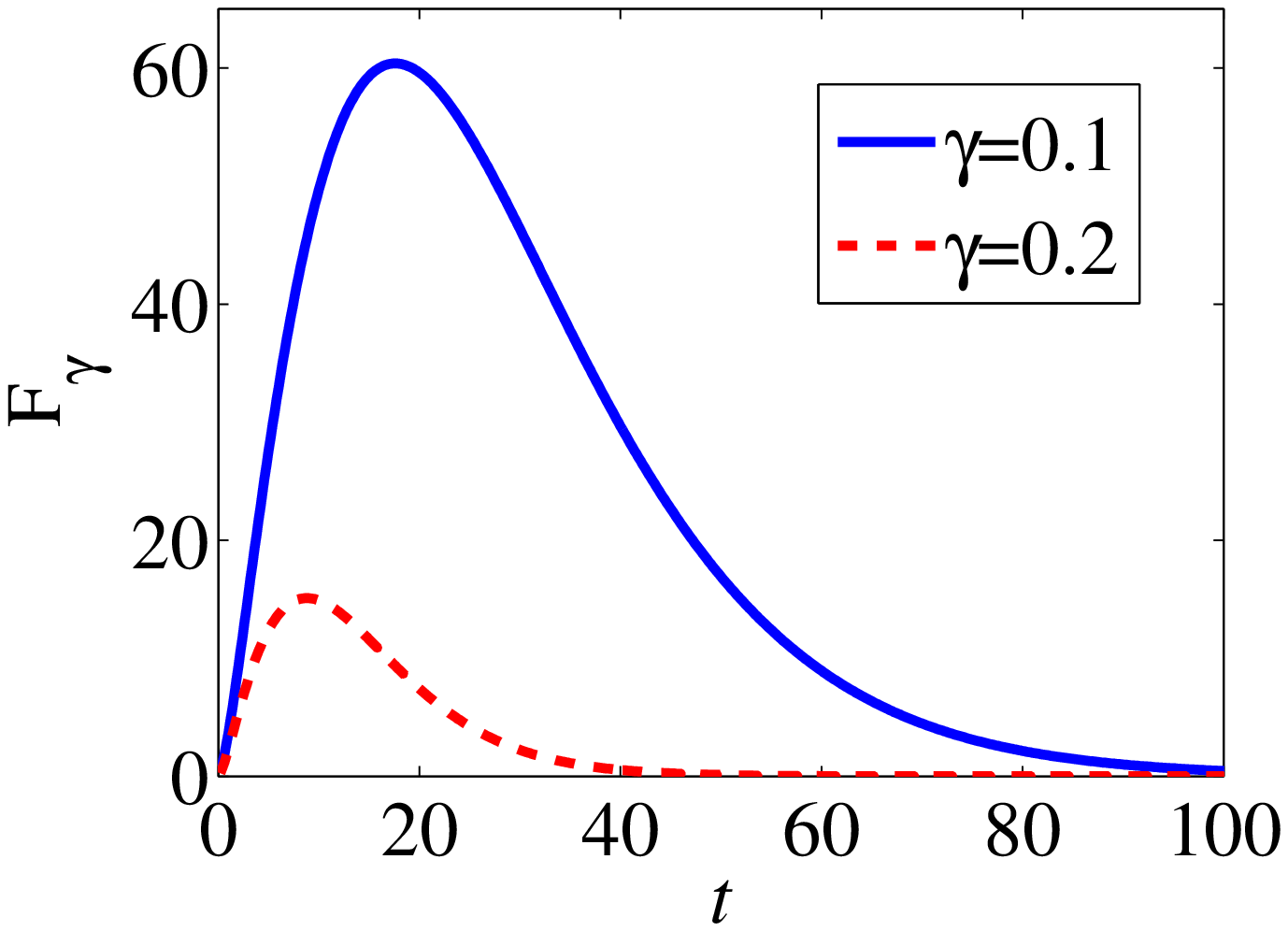}
\includegraphics[width=1.5in]{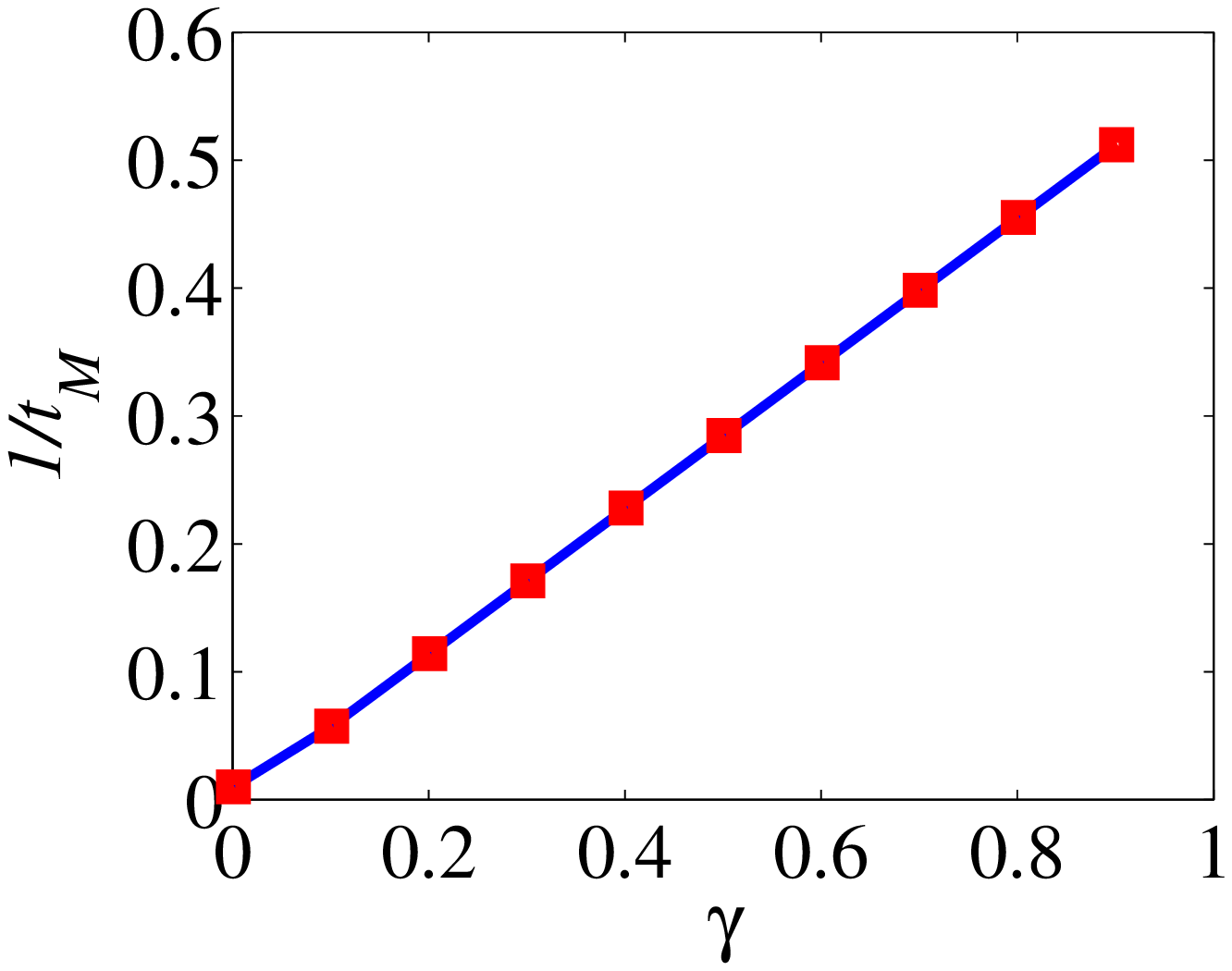}
\hspace{2.2cm}\caption{Left panel: The evolution of
$F_{\gamma}$ with respect to time $t$ under different decoherence strengths.
A large decoherence rate makes the precision of estimation decrease.
Right panel: The variation of $1/t_{M}$ in term of $\gamma$. They are linearly related.
Here $t_{M}$ denotes the time at which $F_{\gamma}$ gets the maximum value.
The parameter is $a=0.5$.} \label{Fs2}
\end{figure}

After the direct diagonalization, the eigenvalues of $\rho(t)$ is given as
\begin{equation}
\begin{array}{llll}
\lambda_{1}=\frac{1}{3} a p(-t)^2,~~
\lambda_{2}=\frac{1}{3} a p(-t)^2 ( p(t)-1), \\
\lambda_{3}= \frac{1}{3} p(-t)^2[a-2 (a+1)p(t)]+1, \\
\lambda_{4}= \frac{1}{3} p(-t)^2[(a+2)p(t)-a],
\end{array}
\end{equation}
with the corresponding eigenvectors
\begin{equation}
\begin{array}{llll}
|1\rangle =(1, 0, 0, 0)^{T},~~|3\rangle =(0, 0, 0, 1)^{T},
\\
|2\rangle =N_{1}(0, -2 e^{i \chi }(\Delta+1), \Gamma, 0)^{T},
\\
|4\rangle =N_{1}(0, -2 e^{i \chi }(\Delta-1), \Gamma, 0)^{T}.
\end{array}
\end{equation}
Here
\begin{equation}
\begin{array}{ll}
\Gamma=2 e^{2 i \chi } \sin ^2(t v)+\cos (2 t v)+1,
\Delta=\sin (\chi ) \sin (2 t v),
\end{array}
\end{equation}
$N_{1}$ is a normalized constant and $A^{T}$ denotes the transposition
of the matrix $A$.

\subsection{Estimation on decoherent rates}
As mentioned above, in this paper we have assumed that the decoherent rates of
the two qubits $A$ and $B$ are equal $\gamma_{A}$=$\gamma_{B}$=$\gamma$. If one considers this
decoherence as a quantum channel, it is reasonable to adopt the two qubits
as a whole to estimate the decoherent parameter $\gamma$ of the environment, as shown in Fig.~\ref{system}(a).
In order to achieve this goal, we adopt the upper mentioned quantum estimation method and focus
on the the dynamics of quantum Fisher information. According to Eq.~(\ref{FisherA}), the quantum Fisher information of $\rho(t)$ with respect to $\gamma$ is expressed as
\begin{equation}
\begin{array}{llll}
\textit{F}_{\gamma}(t)= \frac{t^2}{3}[\frac{4 (a^2 -a +1)}{-2 a p(t)+a-2 p(t)+
3 p(t)^2}+\frac{(a+2)^2}{a p(t)-a+2 p(t)}+\frac{a}{p(t)-1}].
\end{array} \label{FisherGamma}
\end{equation}
It's obvious that $\textit{F}_{\gamma}(t)$ is independent of parameters $v$ and
$\chi$. This can be easily understood by the following fact: only the decay process from the excited state
$|e\rangle$ to the ground state $|g\rangle$ contributes to $\textit{F}_{\gamma}(t)$. This process is independent of their interaction and initial coherence $\exp(i \chi)$. Note also that the eigenvectors $|k\rangle$ are independent of $\gamma$, so $\textit{F}_{\gamma}(t)$ only has a classical part according to Eq.~(\ref{FisherA}). Our results indicate that from the perspective of quantum Fisher information,
this entangled mixed state $\rho(t)$ does not show any quantum coherence.
The relationship between quantum Fisher information and entanglement
will be studied in future.

In addition, it would be interesting to consider the asymptotic limits of $\textit{F}_{\gamma}(t)$.
On the one hand, in the limit $\gamma \rightarrow 0$, that the system approaches the unitary evolution,
\begin{equation}
\begin{array}{llll}
\textit{F}_{\gamma}(t)\propto \frac{t^2}{\exp(\gamma t)-1}.
\end{array} \label{Fgammaa}
\end{equation}
In this case, the state preserves its phase coherence and QFI is an increase function at the initial time evolution. On the other hand,
\begin{equation}
\begin{array}{llll}
\textit{F}_{\gamma}(t) \propto \exp(-2\gamma t)
\end{array}
\end{equation}
in the completely decoherent limit $\gamma \rightarrow \infty$. In this case,
the state loses its coherence and QFI is exponential decay.
As $\frac{t^2}{\exp(\gamma t)-1}$ is an increasing function of time and $\exp(-2\gamma t)$ is a decreasing one, $F_{\gamma}$ must have a maximum at a finite time.

\begin{figure}[!tb]
\centering
\includegraphics[width=2.0in]{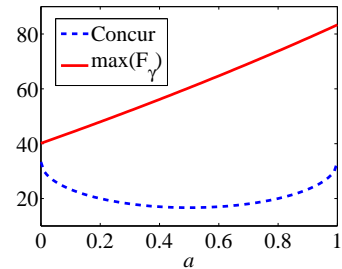}
\hspace{2.2cm}\caption{The variation of the maximum value of $F_{\gamma}(t)$
with respect to $a$. The curve of concurrence
has been amplified by 50 times for clarity. The parameter is $\gamma=0.1$.} \label{Fs7}
\end{figure}

We also adopt the numerical simulations to display other behaviors of quantum Fisher
information. In Fig.~\ref{Fs1}, we show the evolution of quantum Fisher
information with respect to time $t$ and parameter $a$ in Eq.~(\ref{init}).
It is clear that $F_{\gamma}(t)$ has a maximum value at a finite time,
and approaches zero in the long time limit. The left panel in Fig.~\ref{Fs2} shows the effect of the decoherent parameter $\gamma$ on
quantum Fisher information. $F_{\gamma}$ drops
considerably with a small increment of $\gamma$.
The decrease of the maximum value of QFI reflects that
the parameter estimation of the open system becomes more
inaccurate.
The similar result has been obtained in Ref.~\cite{Berrada}. Moreover,
we also found that $1/t_{M}$ is linearly related to $\gamma$, i.e., $\gamma t_{M}= Const.$,
as shown in the right panel of Fig.~\ref{Fs2}.
Here $t_{M}$ denotes the time at which $F_{\gamma}$ gets the maximum value.

As the maximum value of QFI implies the largest precision to estimate $\gamma$, in
Fig.~\ref{Fs7} we plot the maximum value of QFI max($F_{\gamma}$) in terms of
$a$. It shows that max($F_{\gamma}$) monotonously increases with respect to $a$.
This relationship can be understood from the following arguments.
The information of $\gamma$ comes from the spontaneous emission of the excited state in the initial state.
For $a=1$, $\rho_{0}=\frac{1}{3}(|ee\rangle \langle ee|+|\psi \rangle \langle \psi|)$, which has
the largest probability in the excited state compared to the other value of $a$. So this state corresponds to
the largest value of max($F_{\gamma}$). With the similar reason, $\rho_{0}=\frac{1}{3}(|gg\rangle \langle gg|+|\psi \rangle \langle \psi|)$ with $a=0$ has the smallest value of max($F_{\gamma}$).

More interestingly, the input state at $a=0$
has the maximum entanglement, which corresponds to the minimum value of
max($F_{\gamma}$). Generally speaking, entanglement is expected to be helpful in estimating the
parameters of a quantum channel \cite{VGio11, Vuletic10}.
However, its belief is not always true.
For example, a two-qubit maximally entangled state only
achieves the best precision estimation in some \textit{limited} range of depolarization \cite{Fischer, Freyberger, Hotta, Bschorr01}.
Based on the dynamics of quantum Fisher information, we also check that parameter estimation is
not enhanced by the entanglement.

\subsection{Estimation on interacting strength}
The two qubits $A$ and $B$ interacts with each other.
Considering the exchanging symmetry between them, it's natural that one chooses
one qubit such as qubit $B$ to estimate the coupling strength $v$,
as shown in Fig.~\ref{system}(b).
Its reduced density matrix $\rho_{B}= \mathrm{Tr}_{A}(\rho)$
\begin{equation}
\rho_{B}=\left(                 
\begin{array}{cc}   
\frac{1}{3} p(-t)\Omega(t)  & 0 \\  
0 & 1-\frac{1}{3} p(-t)\Omega(t) \\  
\end{array}
\right). \label{redu}                 
\end{equation}

\begin{figure}[!tb]
\centering
\includegraphics[width=2.0in]{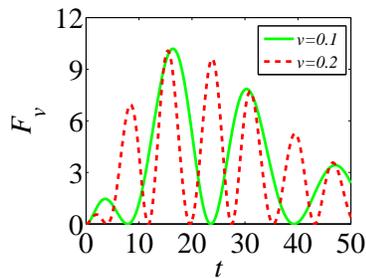}
\hspace{2.2cm}
\caption{The effect of interacting parameter $v$ on the variation of $F_{v}$
with respect to time $t$. $F_{v}(t)$ shows oscillation behavior.
Other parameters are $\gamma=0.1$, $a=0.8$ and $\chi=0.5$.}
\label{Fs4}
\end{figure}

\begin{figure}[!tb]
\centering
\includegraphics[width=1.5in]{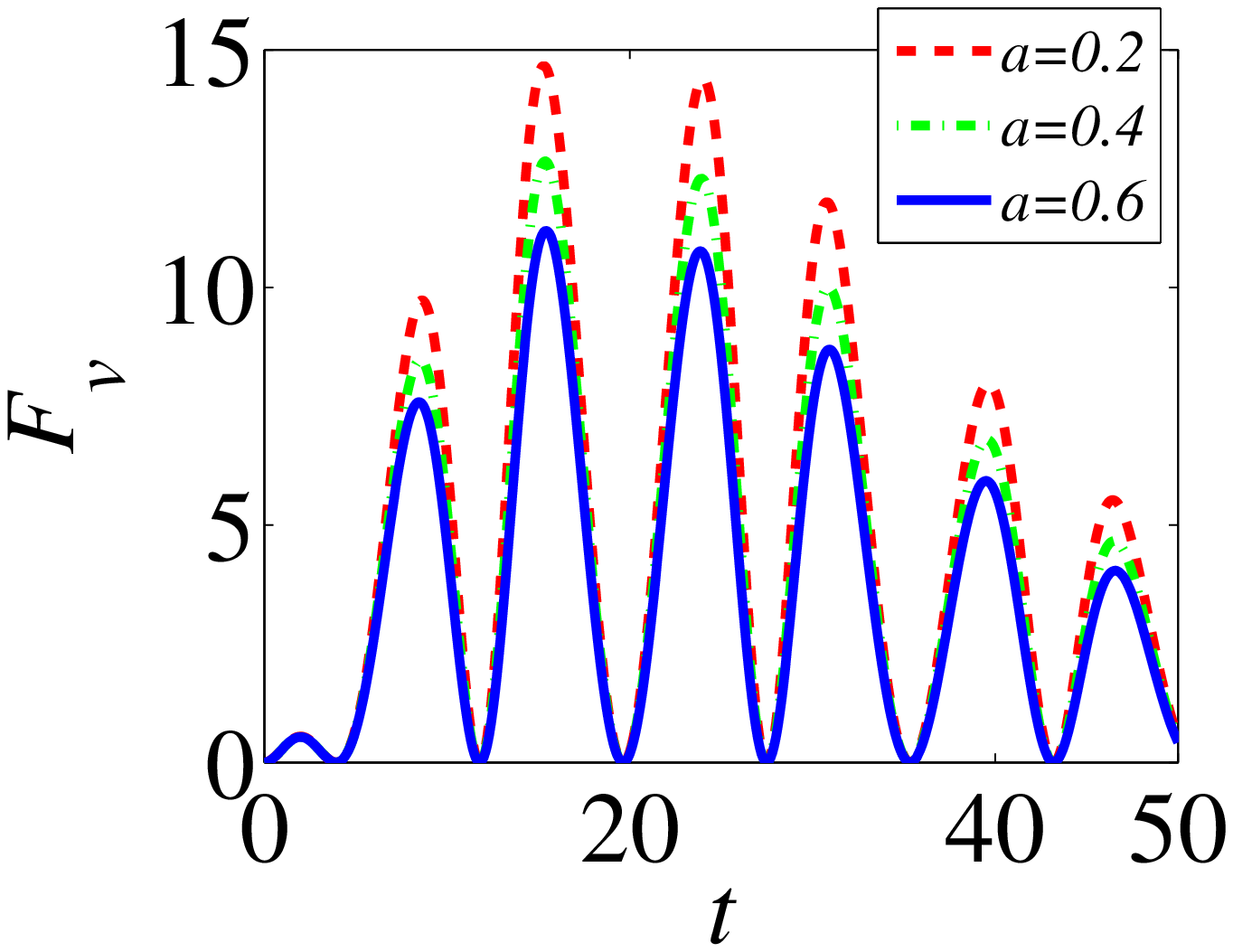}
\includegraphics[width=1.5in]{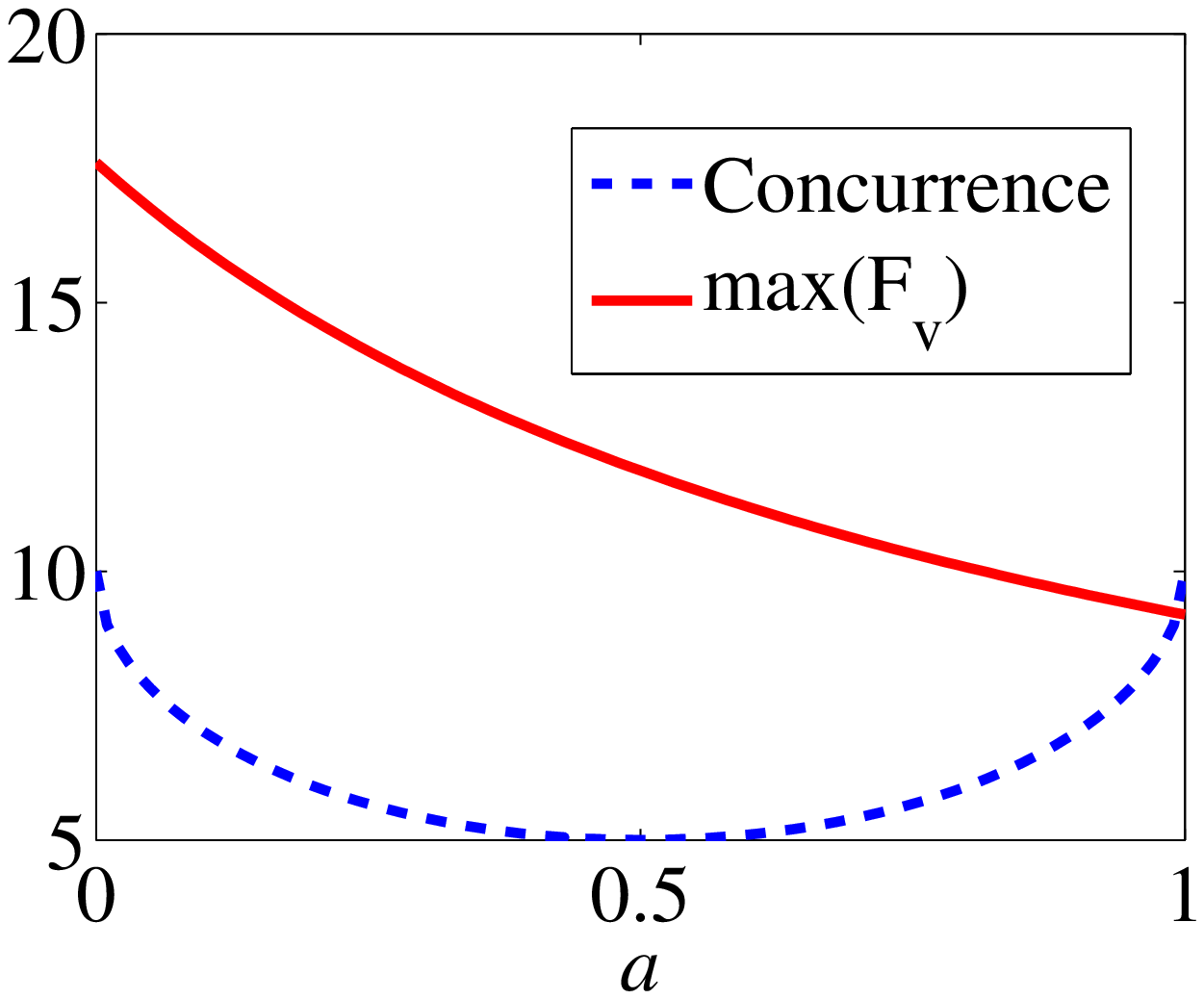}
\hspace{2.2cm}\caption{Left panel: The evolution of $F_{v}(t)$. Different value of
$a$ only affect the amplitude of $F_{v}(t)$. Right panel: The variation
of the maximum value of $F_{v}(t)$ with respect to $a$. The curve of concurrence
has been amplified by 15 times for clarity. Other parameters are
$\gamma=0.1$, $v=0.2$, $\chi=0.5$} \label{Fs5}
\end{figure}

Adopting a similar procedure as above, the QFI of $\rho_{B}$ with respect to
the coupling strength $v$ is
\begin{equation}
F_{v}(t)= \frac{[2 t \sin(\chi ) \cos(2vt)]^2}{\Omega(t)[3 p(t)-\Omega(t)]}
\label{Fisherv}
\end{equation}
with $\Omega(t)=1+a+\sin(2vt) \sin(\chi)$.
In contrast to the two qubits Fisher information Eq.~(\ref{FisherGamma}), here
the initial coherence between $A$ and $B$ plays a role.
For example, $F_{v}(t)=0$ if $\chi$=0 and $\pi$, while
$F_{v}(t)>0$ for the other values of $\chi$.

Eq.~(\ref{Fisherv}) also displays that the information of the interaction parameter $v$ is embedded
in the evolution of $F_{v}(t)$.
The variation of $F_{v}(t)$ in terms of time is shown in Fig.~\ref{Fs4}.
It displays the oscillating behavior and has multi-peaks, which is distinct from the
single-peak of $F_{\gamma}(t)$ as shown in Fig.~(\ref{Fs2}).

We also study the effect of the initial parameter $a$ on quantum Fisher information $F_{v}(t)$,
as an example shown in Fig.~\ref{Fs5}.
With different values of $a$, the amplitudes of $F_{v}(t)$
are changed. The maximum value of $F_{v}(t)$ is a decreasing function
of $a$. Compared to Fig.~\ref{Fs7}, in this figure the input state at $a=1$
has the maximum entanglement, but it corresponds to the minimum value of
max($F_{\gamma}$).

\section{Conclusion}
In conclusion, we have analyzed the dynamics of quantum Fisher information,
which is related to Cram$\acute{e}$r-Rao inequality in quantum estimation
theory, for two initially entangled qubits under decoherence.
We have observed that for a special class of $X$-state, its quantum Fisher information
$F_{\gamma}$ with respect to the decoherent parameter only has the classical part.
Moreover, the dynamical evolution of quantum Fisher
information shows $\frac{t^2}{\exp(\gamma t)-1}$ and $\exp(-2\gamma t)$ in the
unitary and completely decoherent limit, respectively. We also study the
estimation of the coupling strength between two qubits by the
dynamics of quantum Fisher information $F_{v}(t)$ for a reduced qubit.
It is oscillation in term of time, which implies the information of the interaction parameter
may be obtained by the Fourier transformation of $F_{v}(t)$. In both cases,
we do not observe that quantum Fisher information is enhanced by the
entanglement of the input state.
The relationship between quantum Fisher information and quantum entanglement
deserves further study.

\section{Acknowledgements}
We thanks the helpful discussion with Dr. Z. H. Wang, S. W. Li, X. Xiao and
X. W. Xu. This work is partially supported by the National Natural Science
Foundation of China (Grant Nos.~11065005 and and 11365006).



\begin{thebibliography}{99}

\bibitem{Tsang13}
M. Tsang, \textit{Phys. Rev. A} \textbf{88}, 021801(R) (2013).


\bibitem{Yu06}
T. Yu and J. H. Eberly, \textit{Phys. Rev. Lett.} \textbf{97}, 140403 (2006).


\bibitem{Nielsen00}
M. A. Nielsen and I. L. Chuang, \textit{Quantum Computation
and Quantum Information} (Cambridge University Press,
Cambridge, U.K., 2000).


\bibitem{Fujiwara01}
A. Fujiwara, \textit{Phys. Rev. A} \textbf{63}, 042304 (2001).


\bibitem{Chiuri04}
A. Chiuri, V. Rosati, G. Vallone, S. P$\acute{a}$dua, H. Imai,
S. Giacomini, C. Macchiavello, and P. Mataloni, \textit{Phys. Rev. Lett.}
\textbf{107}, 253602 (2011).


\bibitem{Fisher} R. A. Fisher, \textit{Proc. Cambridge Philos. Soc.} \textbf{22}, 700
(1925).

\bibitem{Holevo} A. S. Holevo,
\textit{Probabilistic and Statistical Aspects of Quantum Theory}
(North-Holland, Amsterdam, 1982).


\bibitem{Caves94}
S. L. Braunstein and C. M. Caves, \textit{Phys. Rev. Lett.}
\textbf{72}, 3439 (1994).



\bibitem{Smerzi09}
 L. Pezz$\acute{e}$ and A. Smerzi, \textit{Phys. Rev. Lett.}
 \textbf{102}, 100401 (2009).



\bibitem{Huelga07}
S. F. Huelga, C. Macchiavello, T. Pellizzari, A. K. Ekert,
M. B. Plenio, and J. I. Cirac, \textit{Phys. Rev. Lett.} \textbf{79}, 3865
(1997).


\bibitem{Escher11}
B. M. Escher, R. L. de Matos Filho, and L. Davidovich, \textit{Nat.
Phys.} \textbf{7}, 406 (2011);
\textit{Braz. J. Phys.} \textbf{41}, 229 (2011).


\bibitem{wangx11}
J. Ma, Y. Huang, X. G. Wang, and C. P. Sun
\textit{Phys. Rev. A} \textbf{84}, 022302 (2011).



\bibitem{zsun10} Z. Sun, J. Ma, X. Lu, and X. G. Wang,
\textit{Phys. Rev. A} \textbf{82}, 022306 (2010).


\bibitem{Berrada} K. Berrada,
\textit{Phys. Rev. A} \textbf{88}, 035806 (2013).








\bibitem{Kok}
P. Kok, S. L. Braunstein and J. P Dowling,
\textit{J. Opt. B}
\textbf{6}, S811 (2004).


\bibitem{Bollinger}
J. J. Bollinger, W. M. Itano, D. J. Wineland, and D. J. Heinzen,
\textit{Phys. Rev. A} \textbf{54}, R4649 (1996).



\bibitem{Zhong13}
W. Zhong, Z. Sun, J. Ma, X. G. Wang, and F. Nori,
\textit{Phys. Rev. A} \textbf{87}, 022337 (2013).


\bibitem{Berrada12}
K. Berrada, S. A. Khalek, and C. H. R. Ooi,
\textit{Phys. Rev. A} \textbf{86}, 033823 (2012).

\bibitem{Berrada13b}
K. Berrada, \textit{Phys. Rev. A} \textbf{88}, 013817 (2013).


\bibitem{Agarwal09} S. Das and G. S. Agarwal,
\textit{J. Phys. B} \textbf{42}, 141003 (2009).


\bibitem{Yu04}
 T. Yu and J. H. Eberly, \textit{Phys. Rev. Lett.}
 \textbf{93}, 140404 (2004).




\bibitem{VGio11}
V. Giovannetti, S. Lloyd, and L.Maccone, \textit{Nat. Photonics} \textbf{5}, 222
(2011).


\bibitem{Vuletic10}
M. H. Schleier-Smith, I. D. Leroux, and V. Vuletic, \textit{Phys. Rev.
Lett.} \textbf{104}, 073604 (2010).



\bibitem{Fischer} D. G. Fischer, H. Mack, M. A. Cirone, and M. Freyberger
\textit{Phys. Rev. A} \textbf{64}, 022309 (2001).

\bibitem{Freyberger} H. Venzl and M. Freyberger,
\textit{Phys. Rev. A} \textbf{75}, 042322 (2007).



\bibitem{Hotta} M. Hotta, T. Karasawa, and M. Ozawa, \textit{J. Phys. A} \textbf{39},
14465 (2006).


\bibitem{Bschorr01}
T. C. Bschorr, D.G. Fischer, and M. Freyberger, \textit{Phys. Lett. A}
\textbf{292}, 15 (2001).


\end{thebibliography}
\end{document}